# Image-based Proof of Work Algorithm for the Incentivization of Blockchain Archival of Interesting Images

Jake Billings

University of Colorado Denver

## ABSTRACT

A new variation of blockchain proof of work algorithm is proposed to incentivize the timely execution of image processing algorithms. A sample image processing algorithm is proposed to determine “interesting” images using analysis of the entropy of pixel subsets within images. The efficacy of the image processing algorithm is examined using two small sets of training and test data. The interesting image algorithm is then integrated into a simplified blockchain mining proof of work algorithm based on Bitcoin. The incentive of cryptocurrency mining is theorized to incentivize the execution of the algorithm and thus the retrieval of images that satisfy a minimum requirement set forth by the interesting image algorithm. The digital storage implications of running an image-based blockchain are then examined mathematically.

# INTRODUCTION

## Blockchain Algorithms

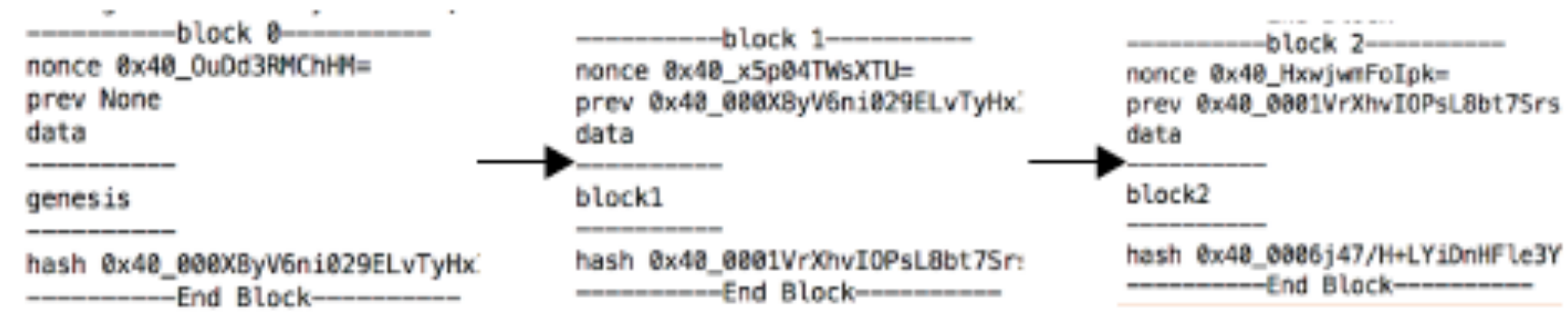

Blockchain algorithms allow distributed networks to gain consensus through distributed proof of work by to solving transaction simultaneity issues. As of the time this paper was written, blockchain consensus algorithms are used widely for transaction agreement in cryptocurrencies, such as Bitcoin and Ethereum. Specifically, blockchains solve conflicts in truth in distributed networks. For instance, consider two transactions withdrawing digital currency from a wallet entered into a distributed network at the same time. This is also known as the double-spend problem. The network must decide on a which transaction occurred first. Miners select transactions to enter into a “block,” and a computationally difficult probabilistic algorithm is used to generate it. Blocks contain the hash of the previous block, the new data to be added to the blockchain, and a random nonce. The first node to publish a block wins because each block depends on the entire previous chain. Nodes executing the computationally intensive Bitcoin “mining” algorithm are incentivized to do so using a reward in the form of the digital currency because miners who find blocks are provided with a reward of Bitcoin (Nakamoto, 3). This incentive combined with growth in value of the currency has resulted in growth of the computing resources expended on Bitcoin hash calculations. Consider the growth of bitcoin mining network from 2009 to 2014. The hash rate, which measures the rate at which miners attempt to find blocks that meet the difficulty requirements of the network, rose from $10^{-3}$ almost $10^{8}$ in the span of five years, (Wang et al. 9). The incentive was large enough to drive hardware innovation in the field of cryptocurrency mining. Engineers developed application specific integrated circuits (ASICs) to run the bitcoin mining algorithm, (O’Dwyer et al. 1).  A majority of the computational power of the Bitcoin network is required to generate random nonces and to hash them with the block data in an attempt to find a block hash that satisfies the rising difficulty requirements of the network. However, blockchain technology could be used to incentivize other activities beyond what is essentially wasting compute power for the mere archival of transaction data. The economic incentive to mine bitcoin led to enormous investment in the development of bitcoin mining technology, (Wang et al. 9). The goal of the research presented here is to apply the economic incentive behind bitcoin mining to other algorithms in order to incentivize the distributed execution of image processing algorithms.

## METHODS

### Decentralized Networking

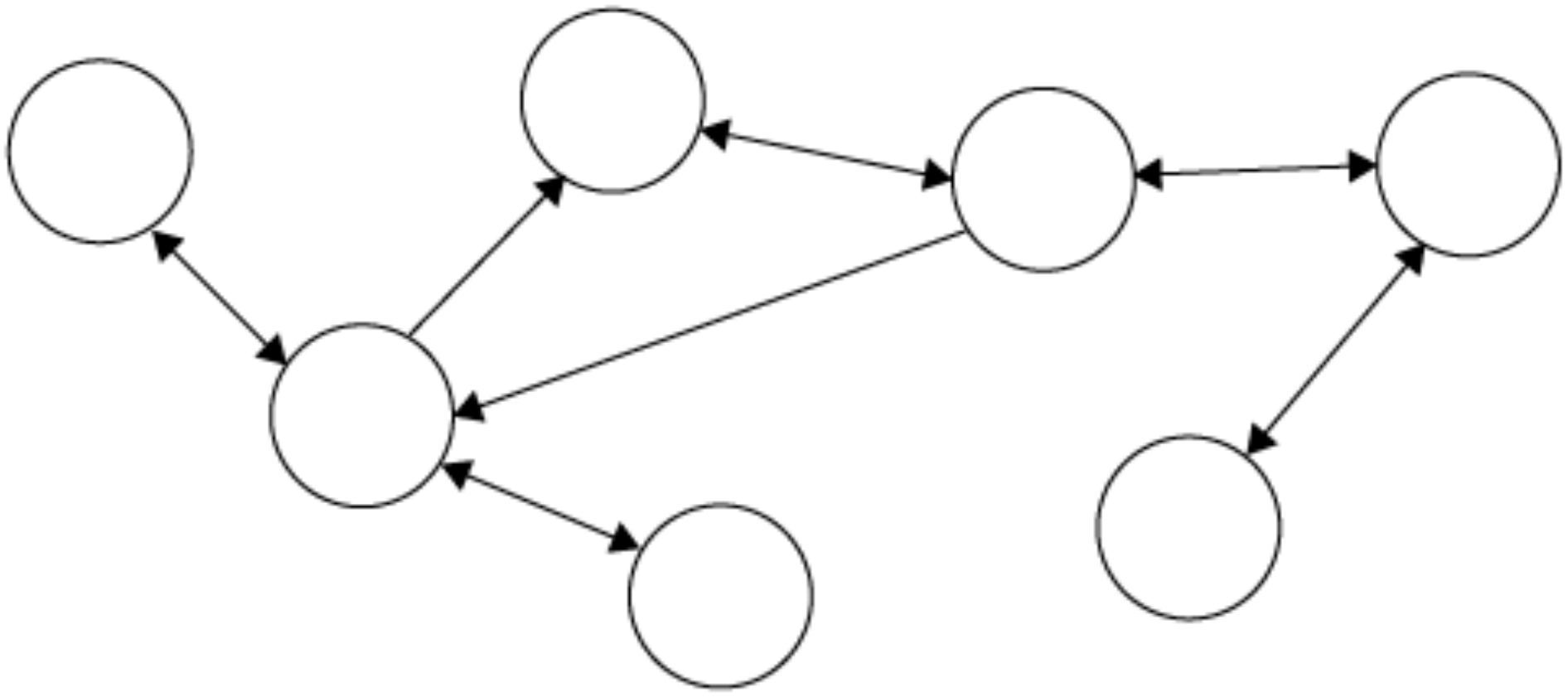

Throughout the rest of this paper, assumptions are made about a decentralized network used to distribute and synchronize the blockchain. It is assumed that any node in the network can publish to all other nodes in the network in the form of a broadcast message. While this paper does not detail the implementation of any such a peer-to-peer network, it is evident from the presence of Bitcoin, Ethereum, and Bittorrent networks that the implementation of peer to peer networks with the desired broadcast functionality is practical. In this type of network, no assumptions can be made about the validity of computation performed by other nodes by any one node in the network because any node could act maliciously to send invalid data to the network as a whole. In fact, it is assumed in each algorithm that other nodes will tamper with or refuse to transmit data arbitrarily or even maliciously. Thus, it is critical that each node validate all data sent to it and that all data must be designed such that tampering is detectable by any node running a correct version of the algorithm. For instance, each node would validate blocks received from the distributed network before committing them to its own stored block chain. The proof of work algorithm guarantees that it is computationally difficult to attack the network with invalid blocks that have valid hashes. Such a network should also incorporate safeguards against spam that could result from malfunctioning nodes or denial of services (DoS) attacks. For instance, each node should validate messages before retransmitting them, and valid messages should be designed to be a small as possible to optimize the network’s overall performance.

## Simplified Blockchain Algorithm

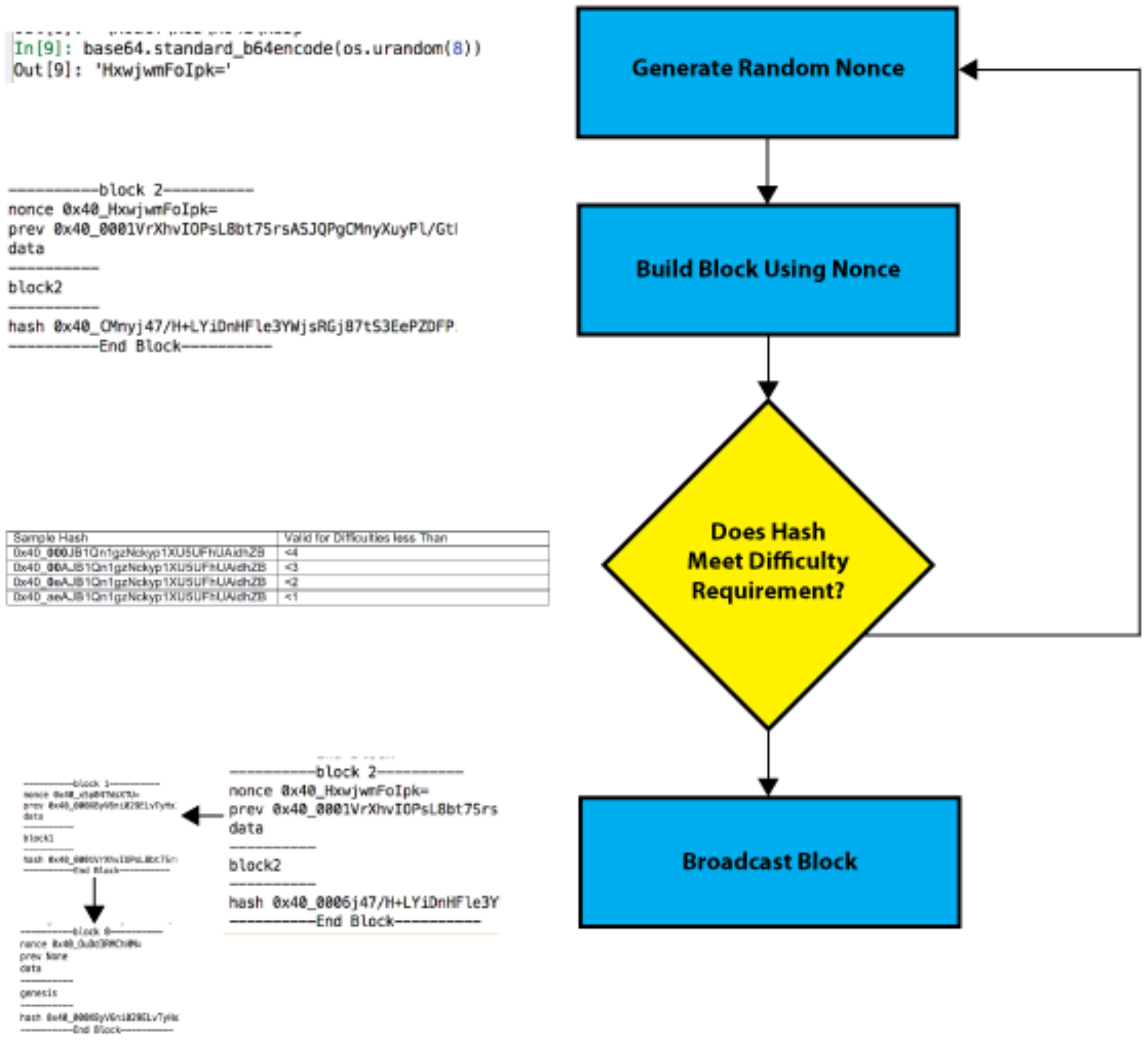

The process of generating a block to add to the blockchain, known as mining, can be performed by any node in the distributed network. To mine a block, a nonce must be found that satisfies the condition that a hash of block data, the previous block's hash, and the nonce must have a given number of leading 0's. It is noteworthy that this is the difference between the bitcoin PoW algorithm and the simplified PoW algorithm found in this paper. In bitcoin, a full less-than comparison is performed between the nonce and an arbitrary difficulty threshold (Nakamoto. 3). This provides bitcoin with more precise control over the difficulty of the mining algorithm as compared to the simplified algorithm. In the simplified algorithm, the difficulty of the block is determined by how many leading figures must be 0s when the hash of the block is encoded using base64. A difficulty order of 2 corresponds to hashes of blocks needing two leading 0s to be considered valid. The hash 0x40_000JB1Qn1gzNckyp1XU5UFhUAidhZB... is valid for a difficulty of 3 or less. (Note that here the base64 encoded string has been prefixed with “0x40_” in order to identify the string as a base64 encoded value). The hash 0x40_00AJB1Qn1gzNckyp1XU5UFhUAidhZB... is valid for a difficulty of 2 or less. In the python implementations seen in this paper, this is known as the DIFFICULTY_ORDER.

*Table 1: Acceptable Hashes for Low Difficulties*

| Sample Hash | Valid for Difficulties less Than |
|---|---|
| 0x40_**000**JB1Qn1gzNckyp1XU5UFhUAidhZB | <4 |
| 0x40_**00**AJB1Qn1gzNckyp1XU5UFhUAidhZB | <3 |
| 0x40_**0**eAJB1Qn1gzNckyp1XU5UFhUAidhZB | <2 |
| 0x40_aeAJB1Qn1gzNckyp1XU5UFhUAidhZB | <1 |

Since hashes including pseudorandom nonces are by nature pseudorandom, and the simplified mining algorithm checks for 0s in the base64 encoded version of the hash, the average and maximum number of hashes required to find a valid hash for an arbitrary difficulty order increases exponentially according to the following:

$$h(\rho) = \alpha^{\rho}$$

Where

$h(\rho)$ = The average number of hashes required to find a valid solution
$\alpha$ = The number of characters used in the encoding
$\rho$ = The arbitrary difficulty order

For this implementation,

$\alpha$ = 64
$\rho$ = 3

Thus,

$$h(\rho) = 64^{\rho}$$
$$h(3) = 64^{3} = 262{,}144$$

Thus, 262,144 hashes or fewer are typically required to mine each block while testing this algorithm. However, the difficulty order can be arbitrarily adjusted by adjusting the variable $\rho$. Changing $\rho$, changes $h$ exponentialy and could be used to maintain a consistent network block generation rate despite computing power that increases exponentially. Theoretically, a production implementation would implement into the blockchain algorithm a system for increasing difficulty based on the current block generation rate such that blocks are generated at intervals that are somewhat regular. This is critical to maintaining a reasonable size of blockchain to be validated and stored by each node.

*Python Implementation of Simplified Algorithm based on Bitcoin*

```
# Mines a block with a given previous block and data
#
# This proof of work (PoW) algorithm is a simplified version of Bitcoin's PoW algorithm.
#
# To mine a block, a nonce must be found that satisfies the condition that a hash of block data, the
previous block's
# hash, and the nonce must have a given number of leading 0's.
#
# previous must be a Block or a None object.
# data must be a string
# difficulty_order must be an integer that refers to the number of leading zeros required in a block's
hash
def mine_block(previous, data, difficulty_order=DIFFICULTY_ORDER):
   # Loop until a block is found. This algorithm could take an arbitrarily long amount of time
   while True:
       # Generate a block based on the previous and the provided data with a random nonce
       b = Block(previous, data, os.urandom(NONCE_SIZE))

       # Fetch the base64 encoded version of the hash. This will be prefixed with the 5 character string
       # "0x40_"
       # This is because the encode module offers both base64 and base32 hashing, and the goal of the
       # module was
       # to provide clear, easy to use encodings. As a result, it is very easy to determine if an encoded
       # string
       # is base64 or base32. 0x40 is 64 in hex. 0x20 is used for base 32.\
       # For instance, a valid hash may look like this: 0x40_000JB1Qn1gzNckyp1XU... Only the
       # "000JB1Qn1gzNckyp1XU" is
       # base64 encoded data. "0x40_" is a prefix.
       hash = b.get_hash_encoded()

       # Assume that the block hash is valid according to the difficulty rules
       valid = True

       # Check in order if each character up to the desired difficulty level is a 0.
       # If it isn't, the nonce does not generate a valid hash, and we must try again with a new nonce
       # on the next iteration.
       for i in range(0, difficulty_order):
           # Add 5 to the starting index because the hash has been prefixed with "0x40_" by the encoding
           # library
           if hash[i+5] is not '0':
               valid = False
               break

       # If the block has been determined to be valid, return it. The block has been mined.
       if valid:
           return b
```

This python implementation will not halt until a block is found and will not give status updates to any users. Theoretical production implementations may consider running the mining algorithm asynchronously and providing users with updates about the number of hashes performed, the rate of hashing, and the status of blocks. Python also presents a serious disadvantage against other miners running implementation of the mining algorithm in other languages since Python is an interpreted language. This example is merely intended to provide a simplistic look into the mining algorithm and how it functions. Early miners in the system could use this implementation until the value of the cryptocurrency incentivizes other developers to write faster versions of the mining algorithm.

*C++ Implementation of Simplified Algorithm based on Bitcoin*

```
Block mineBlock(std::string previous_hash, std::string data, unsigned long
height) {
    while (true) {
        std::string nonce = generateNonce();
        Block b = Block(previous_hash, data, nonce, height);
        const char* chash = b.getHashEncoded().c_str();

        bool valid = true;
        for (unsigned int i=0; i<DIFFICULTY_ORDER; i++) {
            if (chash[i] != '0') {
                valid =false;
                break;
            }
        }
        if (valid) {
            return b;
        }
    }
}
```

A C++ implementation of the mining algorithm similar to the one above will vastly outperform the previous python implementation as C++ runs natively on CPUs as opposed to relying on the overhead of an interpreter. Theoretically, an OpenCL or CUDA implementation could outperform the C++ implementation by running on many specialized graphics cores simultaneously. A GPU implementation has not been included in this paper. However, strong enough economic incentivization from rising cryptocurrency value would theoretically eventually drive miners to develop the algorithm for themselves.

*Python Implementation of Mining 10 Blocks*

```python
import blockchain_like_btc as blockchain

# Print welcome message
print 'Mining at difficulty order %s, which corresponds to a difficulty of %s hashes' %
(blockchain.DIFFICULTY_ORDER, blockchain.DIFFICULTY)

# Mine and print the genesis block, which has no previous block and no information
chain = [blockchain.mine_block(None, 'genesis')]
print chain[0].get_string()

# Mine 10 blocks to demonstrate the working blockchain
for i in range(1, 10):
   b = blockchain.mine_block(chain[i-1], 'block'+str(i))
   chain.append(b)
   print b.get_string()
```

*Sample Output of the Mining of 10 Blocks*

```
Mining at difficulty order 3, which corresponds to a difficulty of 262144.0 hashes
----------block 0----------
nonce 0x40_OuDd3RMChHM=
prev None
data
----------
genesis
----------
hash 0x40_000X8yV6ni029ELvTyHxIYOcWaIcBMSm+S6Cugq0Sh+gjnrM+XKIflCIYP3oV383ipKLmvyxSlcYd1uJESOH7w==
----------End Block----------
----------block 1----------
nonce 0x40_x5p04TWsXTU=
prev 0x40_000X8yV6ni029ELvTyHxIYOcWaIcBMSm+S6Cugq0Sh+gjnrM+XKIflCIYP3oV383ipKLmvyxSlcYd1uJESOH7w==
data
----------
block1
----------
hash 0x40_0001VrXhvIOPsL8bt7SrsASJQPgCMnyXuyPl/GtK4E12Fe7wRwVPJr7smx6ct9Mv1eU7JSHqJQg6ycyenxf/Dg==
----------End Block----------
----------block 2----------
nonce 0x40_HxwjwmFoIpk=
prev 0x40_0001VrXhvIOPsL8bt7SrsASJQPgCMnyXuyPl/GtK4E12Fe7wRwVPJr7smx6ct9Mv1eU7JSHqJQg6ycyenxf/Dg==
data
----------
block2
----------
hash 0x40_0006j47/H+LYiDnHFle3YWjsRGj87tS3EePZDFPi5K8E5IF4lDloGNTg/tZ7KVOKMO7lI+71S/O4htSyhoj/Jw==
----------End Block----------
```

Here, three blocks are mined using the simplified algorithm and a difficulty order of 3. As a result, three blocks are generated that meet this complexity. Three blocks were mined in under five minutes on a MacBook Pro.

## Feature of an Algorithm for the Identification of “Interesting” Images

In order to examine the feasibility of distributed image processing, an algorithm was developed to determine “interesting” images. This algorithm was integrated with the bitcoin blockchain algorithm to incentivize the discovery of images that the algorithm deems “interesting.” Examples of interesting images and uninteresting images were split into training and test datasets. The training set was used to develop an algorithm for classification of images, and test set was used to mitigate risk of over-fitting. Interesting images should have visual structure to humans viewing it. The image should have some sort of meaning if it is deemed by an algorithm to be “interesting.” Images containing one static color or random noise are not interesting to a human observer. Human observers should be able to distinguish shapes and patterns in images they are interested in. Images where humans cannot detect patterns are fundamentally uninteresting. Consider the following subset of examples from the training data set:

| Interesting Images | Uninteresting Images |
|---|---|
| 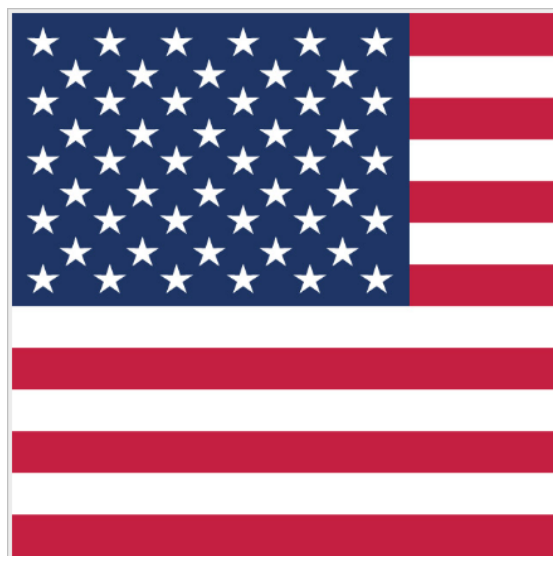 | 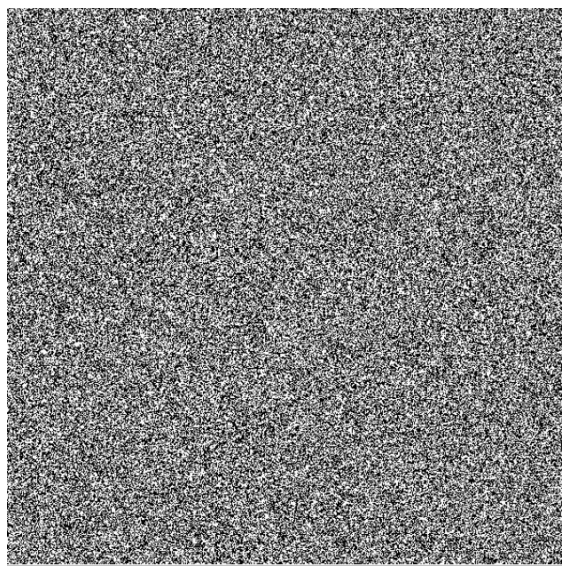 |
| 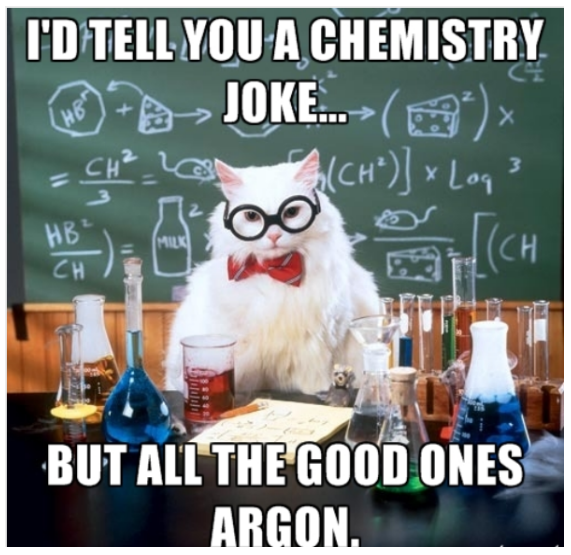 |  |
| 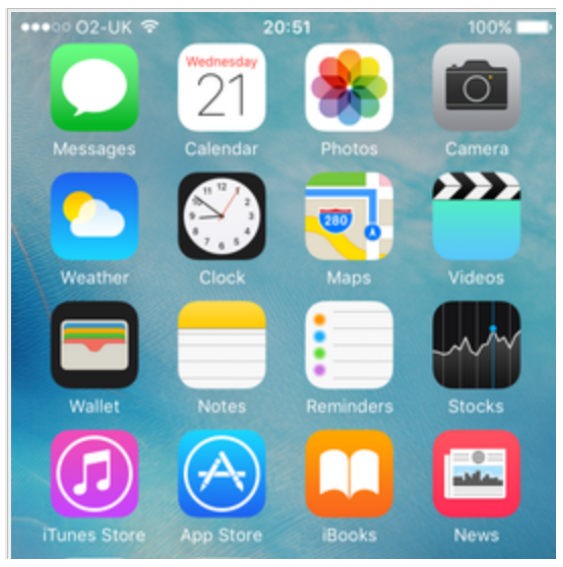 | 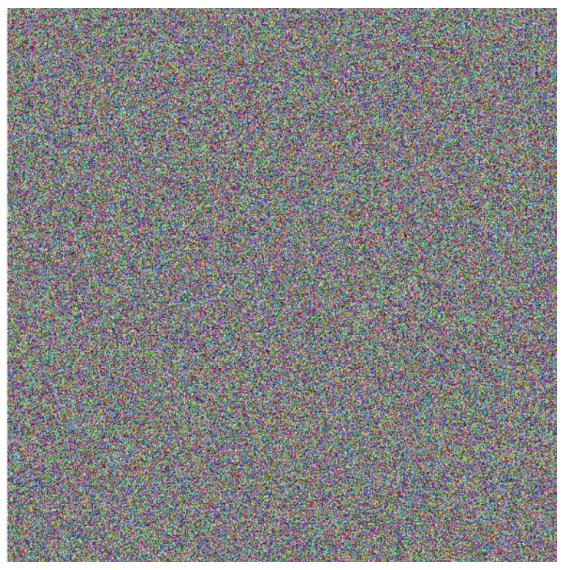 |

## Using Entropy for an Algorithm for the Identification of "Interesting" Images

The final algorithm used by this paper to identify interesting images is uses entropy as the main factor for determining if an image will be interesting. The algorithm uses the summation of the second degree entropy (entropy of the entropy) of images to determine the amount of structure they contain.

First, consider the Shannon entropy of a signal:

$$H(X) = -\sum_{i=1}^{n} P(x_i) \log_b P(x_i)$$

Where

P = the probability of receiving a signal value
$x_i$ = the index of information within the signal
b = the base of the encoding of the signal
n = the distance into the signal to analyze entropy

For two coin flips,

P = 1/b = 1/2
b = 2
n = 2

Thus,

$$H(X) = -\sum_{i=1}^{2} 0.5 \log_2 0.5 = 1$$

In order to process entropy in images, images must be converted into a signal function. The signal is then used to estimate probabilities and ranges algorithmically. As a result, a kernel can be developed to process entropy for the entire image or for small neighborhoods within images resulting in maps of entropy throughout the image.

*Python Implementation of Entropy Calculation of a Signal*

Python signal analysis for processing entropy in images is based on Johannes Maucher's public coursework.

```
# Credit to Johannes Maucher
# https://www.hdm-stuttgart.de/~maucher/ipnotebooks/MMcodecs/01basicfunctions/nb01entropyCalculation.ipynb
def entropy(signal):
  lensig = signal.size
  symset = list(set(signal))
  numsym = len(symset)
  propab = [np.size(signal[signal == i]) / (1.0 * lensig) for i in symset]
  ent = np.sum([p * np.log2(1.0 / p) for p in propab])
  return ent
```

First, a kernel selects 3x3 subsets of the larger image. The entropy of this "neighborhood" will be calculated and applied to the center pixel. This process is repeated for every pixel.

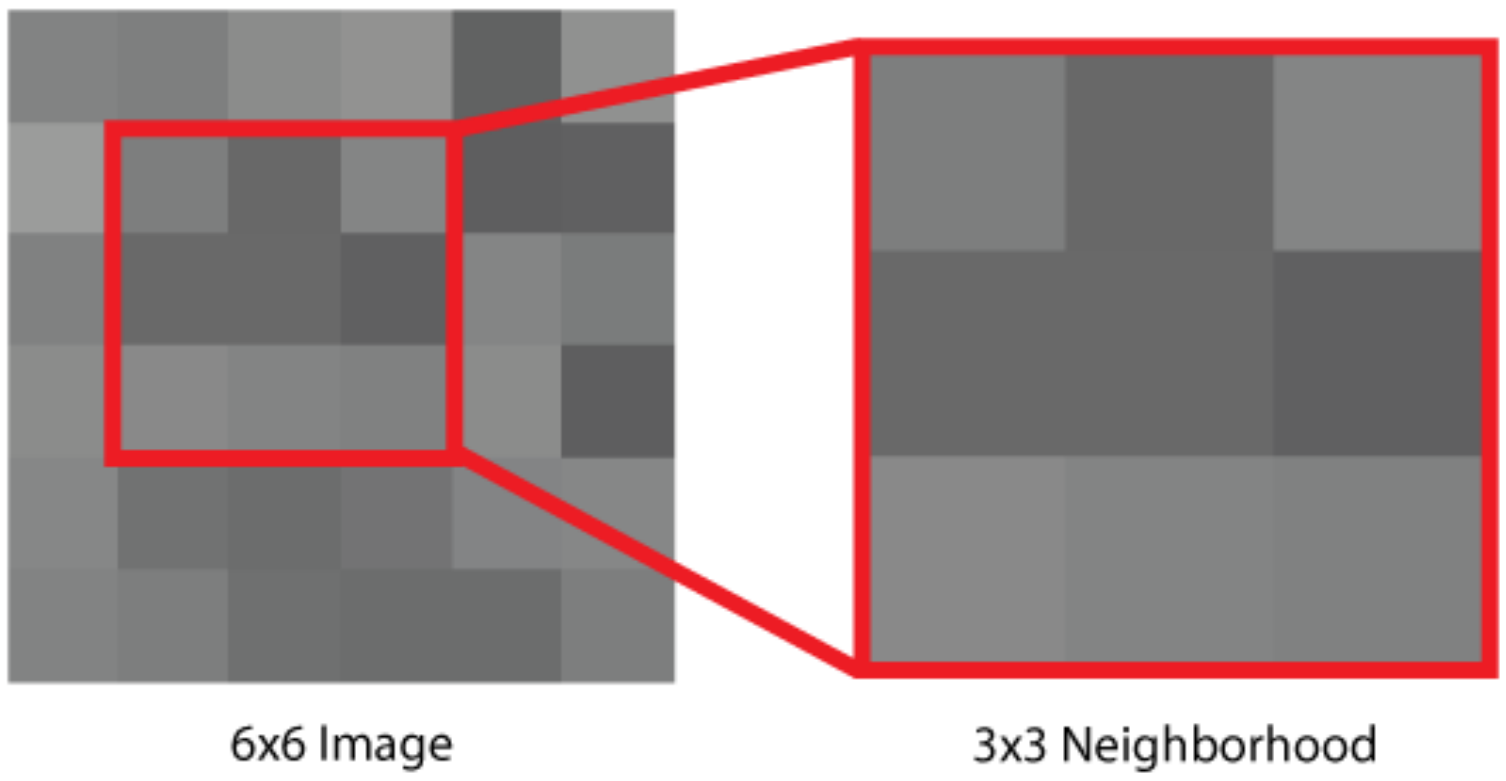

The neighborhood is then flattened into a signal that can be represented as a row of pixels.

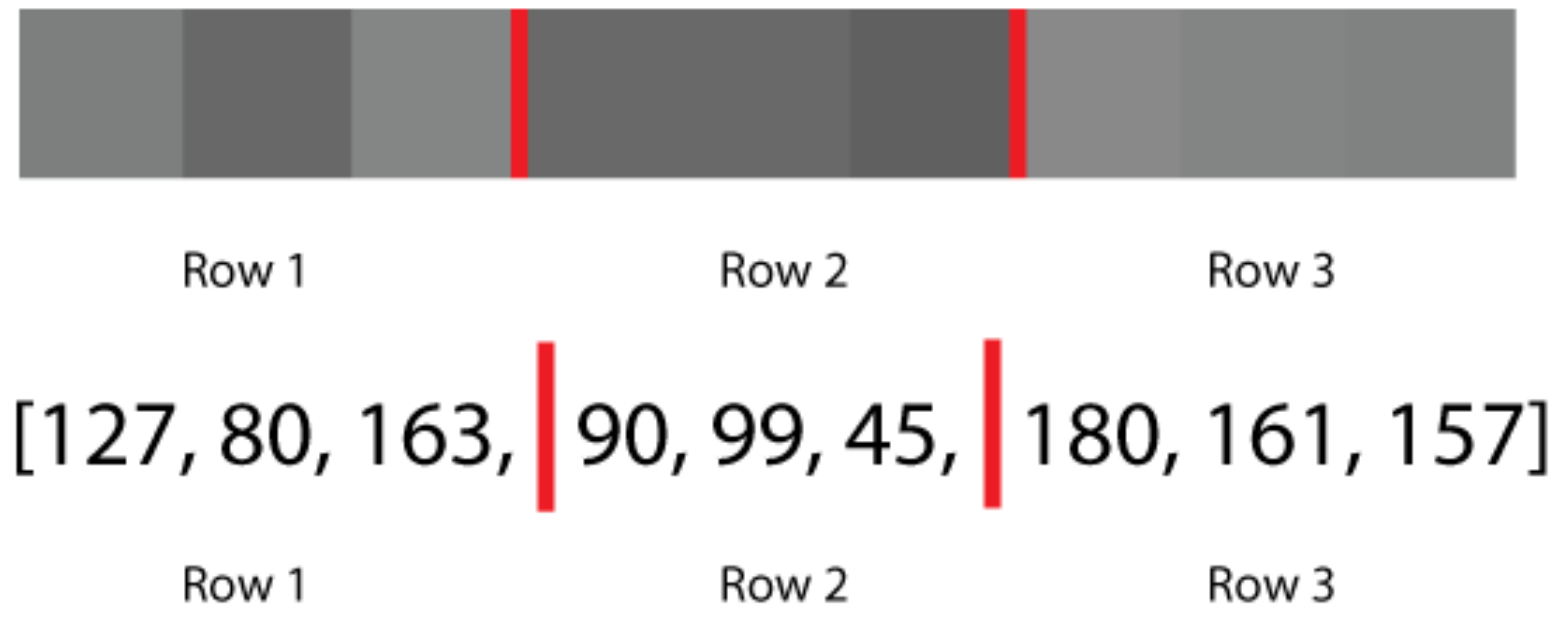

*Python Implementation of Calculation of the Neighborhood Entropy of a Numpy Matrix*

Python signal analysis for processing entropy in images is based on Johannes Maucher's public coursework.

```
# Credit to Johannes Maucher
# https://www.hdm-
stuttgart.de/~maucher/ipnotebooks/MMcodecs/01basicfunctions/nb01entropyCalculation.ipynb
def entropy_matrix(greyIm, N = 3):
  S = greyIm.shape
  E = np.array(greyIm)
  for row in range(S[0]):
     for col in range(S[1]):
        Lx = np.max([0, col - N])
        Ux = np.min([S[1], col + N])
        Ly = np.max([0, row - N])
        Uy = np.min([S[0], row + N])
        region = greyIm[Ly:Uy, Lx:Ux].flatten()
        E[row, col] = entropy(region)

  return E
```

*Python Implementation Conversion of Image to Greyscale Numpy Matrix*

```
# Returns a matrix representing the entropy of a color image by converting it to a
# greyscale image matrix and then analyzing the entropy within
def entropy_matrix_image(colorIm):
   greyIm = colorIm.convert('L')
   colorIm = np.array(colorIm)
   greyIm = np.array(greyIm)

   return entropy_matrix(greyIm)
```

Entropy of color images is calculated by first converting them to greyscale images. This conversion is necessary to acquire a single intensity value for each pixel to enter into a matrix in order to analyze neighborhood entropy.

*Python Implementation of the Summation of Second-degree Entropy in an Image Matrix*

```
def complexity_score(image):
  return np.sum(entropy_matrix(entropy_matrix_image(
    image
  )))
```

*np.sum* is equivalent to the following:

$$\sum_{i=1}^{I}\sum_{j=1}^{J} a_{ij}$$

Where

$$\text{I} = \text{image width}$$
$$\text{J} = \text{image height}$$
$$a_{ij} = \text{Matrix value at I, J}$$

The python function complexity_score returns how “interesting” an image appears to be. The efficacy of this metric is analyzed in a later section.

*Python Implementation of Scaling and Scoring of Images*

```
def is_interesting(image, bottom_threshold=500, width=80, height=80):
    image.resize((width, height), Image.ANTIALIAS)
    score = complexity_score(image)

    return bottom_threshold < score
```

In order to standardize the measure of entropy of each image and to reduce the compute power needed for each iteration, images were scaled to 80px squares using the PIL resize function with antialiasing enabled. Scores were calculated as shown above in the complexity_score() function. Images with a second degree entropy summation greater than bottom_threshold were marked as interesting by the algorithm. This method proved to be fairly effective at separating “interesting” images from “boring” or “noninteresting” images.

Entropy and Second-Degree Entropy of a Sample Image

*Intensities of First and Second Degree Entropy*

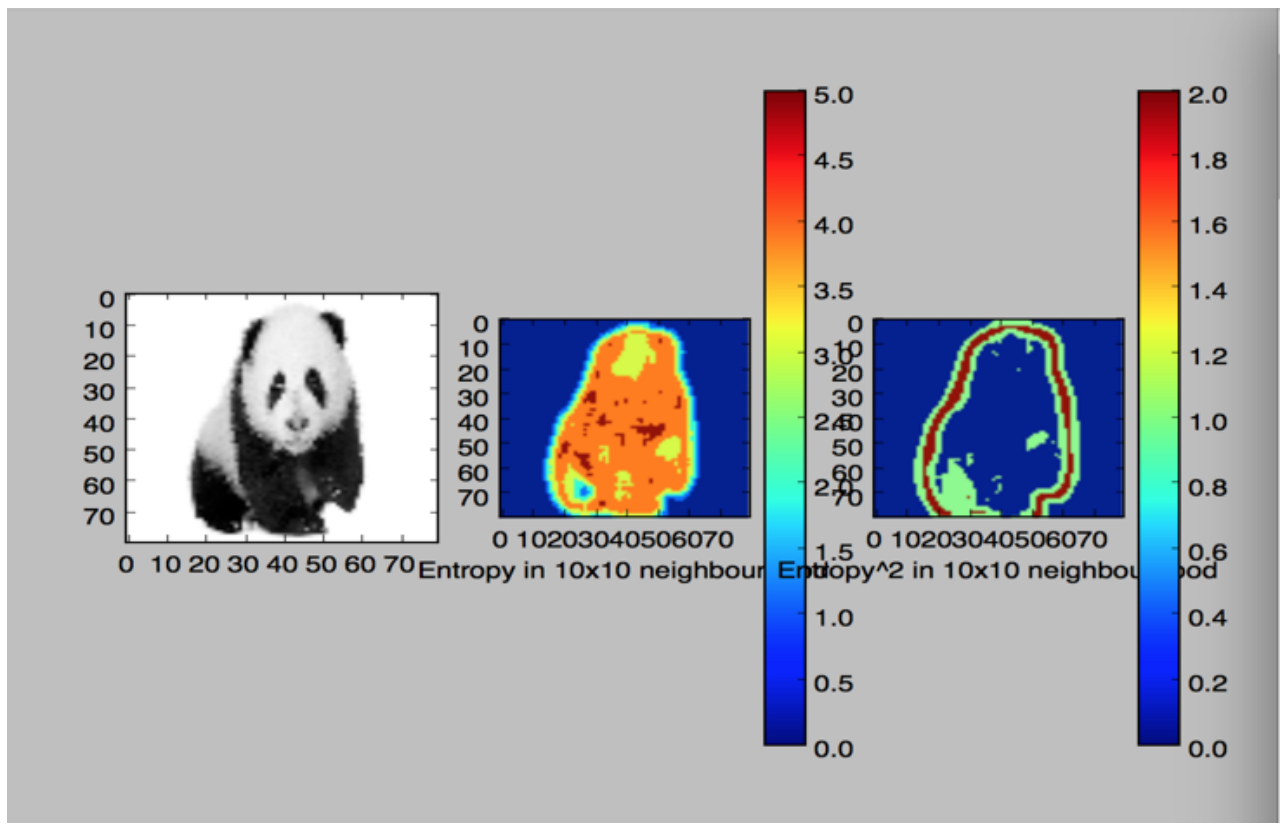

Packages from the Anacoda Python distribution were used to export the steps of the analysis of the image. On the left is the scaled and grey-scaled version of the input image. The middle image shows the entropy map of that image. The rightmost image shows the second-degree entropy of the image.

| | | |
|---|---|---|
| Scaled and Grey-scaled Image | 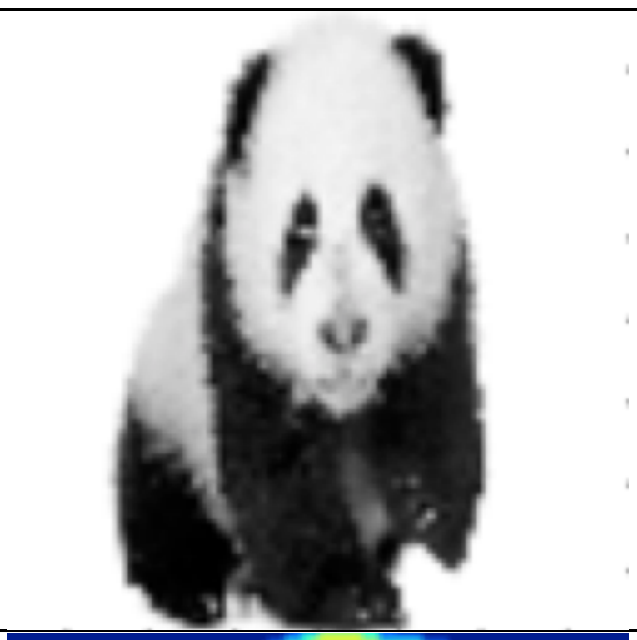 | This is the result of scaling a public domain 400x300 image of a panda. Scaling images standardizes the final result of the entropy summation and also reduces the required compute power. |
| Entropy of Scaled and Grey-scaled Image | 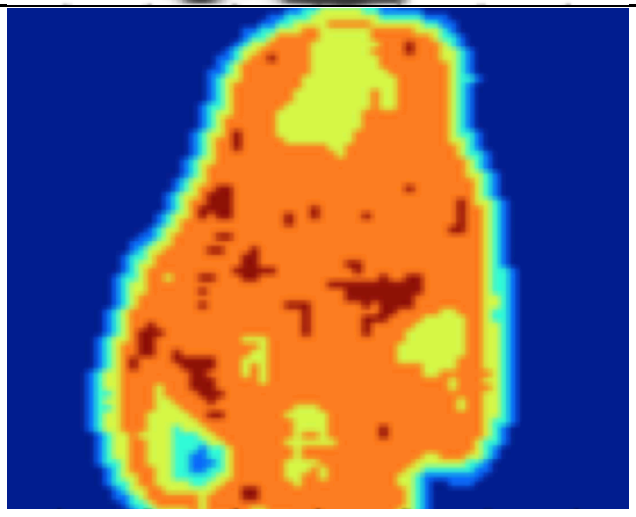 | This map of entropy shows where the kernel-based Sharron entropy of the image is highest. The first-degree entropy does well as distinguishing objects from background. |
| *Second-Degree Entropy of Scaled and Grey-scaled Image* | 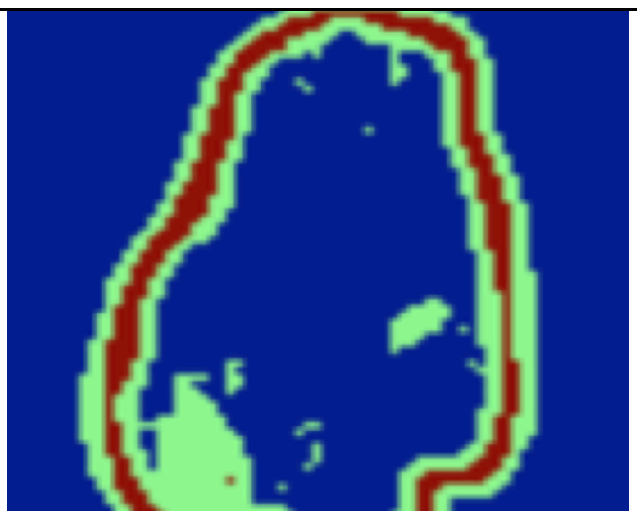 | This map of entropy shows where the kernel-based Sharron entropy of the entropy of the image is highest. The second-degree entropy works almost as an edge detector. |

## Integration of Entropy-Based Image Processing into Simplified Blockchain

The processing of images to determine if they are “interesting” is not a proof of work. As a result, if image efficacy were the only metric for adding blocks, the blockchain could easily be flooded with thousands of blocks. This would waste space on the blockchain and make it infeasible to store. Instead, the image interesting image algorithm could be integrated into the existing Bitcoin algorithm. This maintains the the arbitrary difficulty capacity of a PoW algorithm while still incentivizing the network to locate interesting images. In this new algorithm, images take the place of randomly generated nonces. When nodes validate new blocks they have mined or received from networked peers, they must validate that each nonce is in fact an image that meets requirements for an “interesting” image. Other image processing and nonce-restraining algorithms could also be employed. For instance, a neural-network-based algorithm could be used instead of an entropy-based algorithm could be used to incentivize the location or generation of images with other desirable qualities.

*Python Implementation of Modifications to the Simplified Bitcoin Mining Algorithm for the Incorporation of Restrained Nonces*

```
def mine_block(previous, data, nonce_source):
  while True:
      nonce = nonce_source.provide_nonce()

      print "trying nonce %s" % encode(nonce)

      b = Block(previous, data, nonce)
      hash = b.get_hash_encoded()

      print "got hash %s" % hash

      valid = True
      for i in range(0, DIFFICULTY_ORDER):
          if hash[i+5] is not '0':
              valid = False
              break

      if valid:
          return b
```

This allows for the incorporation of arbitrary nonce generation for the mining of blocks. The following class is in example of a generator that could be used with this function, and it uses the image processing algorithm to select these nonces.

*Python Implementation of a Class for the Validation and Random Access of Potential Nonce Images from a Folder*

```python
import os
import random

sys_random = random.SystemRandom()


from PIL import Image

import images
import cStringIO


class NonceSource:
  def __init__(self, root):
      self.interesting_filenames = os.listdir(root)
      self.interesting_images = []
      for name in self.interesting_filenames:
          self.interesting_images.append(Image.open(os.path.join(root,name)))

  def provide_image(self, limit=10, width=80, height=80):
      for i in range(0,limit):
          image = sys_random\
              .choice(self.interesting_images)\
              .resize((width,height))\
              .convert('RGB')

          if images.is_interesting(image):
              return image

  def provide_nonce(self, limit=10, width=80, height=80):
      buffer = cStringIO.StringIO()
      img = self.provide_image(limit, width, height)
      img.save(buffer, format="JPEG")
      return buffer.getvalue()
```

*Example Genesis Block Mined using an Image Nonce Encoded with Base64*

```
----------block 0----------
nonce 0x40_/9j/4AAQSkZJRgABAQAAAQABAAD/2wBDAAgGBgcGBQgHBwcJCQgKDBQNDAsLDBkSEw8UHRofHh0aHBwgJC4nICIsIxwcKDcpLDAxNDQ0Hyc5PTgyPC4zNDL/2wBDAQkJCQwLDBgNDRgyIRwhMjIyMjIyMjIyMjIyMjIyMjIyMjIyMj
prev None
data
----------
genesis
----------
hash 0x20_dkcus6iez4cuyyq642hbuwf7fw6hphsh25vvdrcketjoqfkg3dhhyi3bjeqjpm7yefkvrdztkbgky7ge6mnm3f2zyuhm456atxwbweq=
----------End Block----------
```

# RESULTS

## Efficacy of Entropy-Based “Interesting” Metric

*Table 2: Algorithm Efficacy of Classification and Test and Training Data*

| TOTAL | | |
|---|---|---|
| | Absolute | Ratio |
| Correct Classifications | 26 | 81.25% |
| Incorrect Classifications | 6 | 18.75% |
| Total Classifications | 32 | 100.00% |

*Table 3: Algorithm Efficacy at Classification of Training Data*

| TRAINING | | |
|---|---|---|
| | Absolute | Ratio |
| Correct Classifications | 12 | 85.71% |
| Incorrect Classifications | 2 | 14.29% |
| Total Classifications | 14 | 100.00% |

*Table 4: Algorithm Efficacy at Classification of Test Data*

| TEST | | |
|---|---|---|
| | Absolute | Ratio |
| Correct Classifications | 14 | 70.00% |
| Incorrect Classifications | 6 | 30.00% |
| Total Classifications | 20 | 100.00% |

These results are based on computation performed by the code in seen the listing *Python Implementation of Image Classification Analysis* and was performed on the results found in Table A1. Both the listing and the table can be found in the appendix.

## Analysis of the Growth of an Integrated Image Processing Blockchain

A blockchain in which each block contains an entire 80px x 80px jpeg image could potentially grow in size much more quickly than a purely text-based blockchain. For instance, the size of the Bitcoin blockchain due to block headers is predicted to grow at a rate of 4.2 MB/year, (Nakamoto, 4). Assuming the worst-case scenario, each image would be stored without compression. A practical image processing blockchain would use compression, which guarantees the actual growth in size of the blockchain will be more favorable than this forecast. However, based on these assumptions, the worst-case scenario for the size of each block is given by the following:

$$s_{block} = 3 \text{ colors} \cdot \frac{1 \text{ byte}}{1 \text{ color}} \cdot \frac{80\text{px} \cdot 80\text{px}}{1 \text{ image}} \cdot \frac{1 \text{ image}}{1 \text{ block}} + \frac{t}{1 \text{ block}} = \frac{19{,}200 \text{ bytes} + t}{1 \text{ block}}$$

Where

$s_{block}$ = the total size of a block in bytes
$t$ = size of the block data in bytes

It can be safely assumed,

$$t < 800 \text{ bytes}$$

Thus,

$$s_{block} < 20{,}000 \text{ bytes}$$

Assuming a rate of generation similar to Bitcoin (Nakamoto),

$$\frac{dn}{dt} \approx \frac{1 \text{ block}}{10 \text{ minutes}}$$

Where

$n$ = the number of blocks

Then,

$$\frac{ds_{chain}}{dt} \approx \frac{dn}{dt} \cdot s_{block}$$

$$\approx \frac{1 \text{ block}}{10 \text{ minutes}} \cdot \frac{60 \text{ minutes}}{1 \text{ hour}} \cdot \frac{24 \text{ hours}}{1 \text{ day}} \cdot \frac{365.25 \text{ days}}{1 \text{ year}} \cdot \frac{20{,}000 \text{ bytes}}{1 \text{ block}}$$

$$= \frac{1{,}051{,}920{,}000 \text{ bytes}}{1 \text{ year}} \approx \frac{1 \text{ gigabytes}}{1 \text{ year}}$$

Where

$s_{chain}$ = size of the chain in bytes

# CONCLUSION

## Efficacy of “Interesting” Image Algorithm based on Entropy

The “interesting” image algorithm proposed in this paper is 81% accurate at classifying images as “interesting” based on training data. Qualitatively, these images appear to a viewer to have structure and to convey information. The algorithm classifies more false negative than false positives. For the application of archival of interesting images, this is more acceptable than a majority false negatives. This is because an algorithm can just fetch more images in order make more attempts to mine a block by submitting an image. Additionally, the algorithm is just efficient enough to be practical to run for a large blockchain when large amounts of compute power are present but difficult enough to justify execution on a distributed system if the compute power would otherwise be going to waste trying random hashes.

## Feasibility of Storage of Blockchain Despite Increased Size

The amount of processing required to validate the blockchain would increase dramatically as compared to the bitcoin blockchain. However, this increase would be proportional. Since the difficulty of the mining adjusts automatically based on network hash rate, this would not decrease the efficacy of the network. Instead of wasting computation on increased hash rate, computation will be spent processing images using the “interesting” image algorithm. The growth of an image-based chain would likely be close to 1GB/year. Based on modern computing equipment, it seems perfectly reasonable to expect nodes participating in the network to store this amount of information. This is especially true if the images in the chain bear some sort of significance, which they will because of the “interesting” image algorithm.

## CITATIONS

Kroll, Joshua A., Ian C. Davey, and Edward W. Felten. "The economics of Bitcoin mining, or Bitcoin in the presence of adversaries." Proceedings of WEIS. Vol. 2013. 2013. <http://www.thebitcoin.fr/wp-content/uploads/2014/01/The-Economics-of-Bitcoin-Mining-or-Bitcoin-in-the-Presence-of-Adversaries.pdf>.

Maucher, Johannes. "Information Theory." Information Theory — Multimedia Codec Excercises 1.0 documentation. Hochscheule Der Medien, n.d. Web. 14 July 2017. <https://www.hdm-stuttgart.de/~maucher/Python/MMCodecs/html/basicFunctions.html>.

Nakamoto, Satoshi. "Bitcoin: A peer-to-peer electronic cash system, 2008." (2012): 1-9. <http://s3.amazonaws.com/academia.edu.documents/32413652/BitCoin_P2P_electronic_cash_system.pdf?AWSAccessKeyId=AKIAIWOWYYGZ2Y53UL3A&Expires=1500047691&Signature=jwnreHG4IbMgmWLTozfq8x8W8hw%3D&response-content-disposition=inline%3B%20filename%3DBitcoin_A_Peer-to-Peer_Electronic_Cash_S.pdf>.

O'dwyer, K.j., and D. Malone. "Bitcoin Mining and Its Energy Footprint." 25th IET Irish Signals & Systems Conference 2014 and 2014 China-Ireland International Conference on Information and Communities Technologies (ISSC 2014/CIICT 2014) (2014): n. pag. Web. 14 July 2017. <http://eprints.maynoothuniversity.ie/6009/1/DM-Bitcoin.pdf>.

Wang, Luqin, and Yong Liu. "Exploring Miner Evolution in Bitcoin Network." Passive and Active Measurement Lecture Notes in Computer Science (2015): 290-302. Web. 14 July 2017. <http://wan.poly.edu/pam2015/papers/23.pdf>.

Wood, Gavin. "Ethereum: A secure decentralised generalised transaction ledger." *Ethereum Project Yellow Paper* 151 (2014). <http://www.cryptopapers.net/papers/ethereum-yellowpaper.pdf>.

# APPENDIX

## Supplementary Python and C++ Code

*Python Implementation of Block Module including Block Class and Related Constants*

```
import hashlib
import os
import math

# Import a custom encoding library
# The encode() function returns a base64 encoded string representing the contents of any
# binary data passed to the function
from encode import encode

# Select the hashing algorithm to be passed to hashlib.new()
# SHA512 was selected due to its low probability of collision even when compared to other hasing
algorithms
# Suggested Value: 'sha512'
HASH_ALGORITHM = 'sha512'

# The amount of random information to pull from os.urandom() in bytes for the nonce of each block
# Suggested Value: 8
NONCE_SIZE = 8


# The difficulty of the block is determined by how many leading figure must be 0s
# A difficulty order of 2 corresponds to hashes of blocks needing 2 leading 0s to be considered valid.
# The hash 0x40_000JB1Qn1gzNckyp1XU5UFhUAidhZB... is valid for a difficulty of 3 or less
# The hash 0x40_00AJB1Qn1gzNckyp1XU5UFhUAidhZB... is valid for a difficulty of 2 or less
# Suggested Value for Demos: 3
# Suggested Value for Production would be based on network hash rate: 3
DIFFICULTY_ORDER = 3

# This is inverse of the probability of the output of an encoded hash being a 0
# Since hashes are encoded in base64, there is a 1/64 probability that any given character in the
encoded hash is a 0
# This is used to estimate the difficulty of mining a block by the number of hashes required to find it.
# Suggested Value: 64
ENCODING_ORDER = 64

# Estimate the difficulty of mining a block in hashes using the encoding order and the difficulty order.
DIFFICULTY = math.pow(ENCODING_ORDER, DIFFICULTY_ORDER)


# The block class is intended to store all associated block data in one place.
# It is not intended to create or 'mine' blocks. For this, use mine_block()
class Block:
  # Instantiate the blobk
  # The block height and block hash are computed from the input values.
  def __init__(self, previous, data, nonce):
      self.previous = previous
      self.data = data
```

```
        self.nonce = nonce

        previous_hash = ''
        if self.previous is not None:
            previous_hash = self.previous.get_hash()
        hasher = hashlib.new(HASH_ALGORITHM)
        hasher.update(previous_hash + self.data + self.nonce)
        self.hash = hasher.digest()

        if self.previous is None:
            self.height = 0
        else:
            self.height = self.previous.get_height() + 1

    def get_previous(self):
        return self.previous

    def get_data(self):
        return self.data

    def get_nonce(self):
        return self.nonce

    def get_nonce_encoded(self):
        return encode(self.nonce)

    def get_hash(self):
        return self.hash

    def get_hash_encoded(self):
        return encode(self.get_hash())

    def get_height(self):
        return self.height

    def get_string(self):
        previous_hash = 'None'
        prev = self.get_previous()
        if prev is not None:
            previous_hash = prev.get_hash_encoded()
        return '----------block %s----------\nnonce %s\nprev %s\ndata\n----------\n%s\n----------\nhash
%s\n----------End Block----------'\
              % (self.get_height(), self.get_nonce_encoded(), previous_hash, self.get_data(),
self.get_hash_encoded())
```

Full Repository: https://github.com/jake-billings/research-blockchain

*C++ Implementation of Block Class and Related Constants*

*Block.h*

```
#ifndef PRACTICE_C_BLOCKCHAIN_BLOCK_H
#define PRACTICE_C_BLOCKCHAIN_BLOCK_H

#include <string>

#define DIFFICULTY_ORDER 3
#define ENCODING_ORDER 64

#define NONCE_SIZE 8

const int DIFFICULTY = DIFFICULTY_ORDER  * ENCODING_ORDER;

class Block {

private:
    std::string previousHash;
    std::string hash;
    std::string data;
    std::string nonce;
    unsigned long height;

public:
    Block(std::string previous_hash, std::string data, std::string nonce,
const unsigned long height);


    const std::string &getPreviousHash() const;
    void setPreviousHash(const std::string &previous_hash);

    const std::string &getHash() const;
    void setHash(const std::string &hash);

    const std::string &getData() const;
    void setData(const std::string &data);

    const std::string &getNonce() const;
    void setNonce(const std::string &nonce);

    unsigned long getHeight() const;
    void setHeight(unsigned long height);

    const std::string getNonceEncoded() const;
    const std::string getHashEncoded() const;
    const std::string getPreviousHashEncoded() const;
};

Block mineBlock(std::string previous_hash, std::string data, unsigned long
height);
void printBlock(Block block);

#endif //PRACTICE_C_BLOCKCHAIN_BLOCK_H
```

*Block.cpp*

```
#include "Block.h"
#include "Crypto.h"
#include "Encode.h"
#include <iostream>
#include <openssl/rand.h>

Block::Block(std::string previousHash, std::string data, std::string nonce,
unsigned long height) {
    this->previousHash = previousHash;
    this->data = data;
    this->nonce = nonce;
    this->height = height;

    this->hash = Crypto::hash(encode(previousHash) + data + encode(nonce));
}

const std::string &Block::getPreviousHash() const {
    return previousHash;
}

void Block::setPreviousHash(const std::string &previousHash) {
    Block::previousHash = previousHash;
}

const std::string &Block::getHash() const {
    return hash;
}

void Block::setHash(const std::string &hash) {
    Block::hash = hash;
}

const std::string &Block::getData() const {
    return data;
}

void Block::setData(const std::string &data) {
    Block::data = data;
}

const std::string &Block::getNonce() const {
    return nonce;
}

void Block::setNonce(const std::string &nonce) {
    Block::nonce = nonce;
}

unsigned long Block::getHeight() const {
    return height;
}
```

```
void Block::setHeight(unsigned long height) {
    Block::height = height;
};

const std::string Block::getNonceEncoded() const {
    return encode(this->getNonce());
}

const std::string Block::getHashEncoded() const {
    return encode(this->getHash());
}

const std::string Block::getPreviousHashEncoded() const {
    return encode(this->getPreviousHash());
}

std::string generateNonce() {
    unsigned char cnonce[NONCE_SIZE];
    RAND_bytes(cnonce, NONCE_SIZE);
    return std::string(cnonce, cnonce + sizeof cnonce / sizeof cnonce[0]);
}

Block mineBlock(std::string previous_hash, std::string data, unsigned long
height) {
    while (true) {
        std::string nonce = generateNonce();
        Block b = Block(previous_hash, data, nonce, height);
        const char* chash = b.getHashEncoded().c_str();

        bool valid = true;
        for (unsigned int i=0; i<DIFFICULTY_ORDER; i++) {
            if (chash[i] != '0') {
                valid =false;
                break;
            }
        }
        if (valid) {
            return b;
        }
    }
}

void printBlock(Block block) {
    printf("----------block %d----------\nnonce %s\nprev %s\ndata\n----------
\n%s\n----------\nhash %s\n----------End Block----------\n",
           (int) block.getHeight(), block.getNonceEncoded().c_str(),
block.getPreviousHashEncoded().c_str(),
           block.getData().c_str(), block.getHashEncoded().c_str());
}
```

Full Repository: https://github.com/jake-billings/practice-c-blockchain

*Python Implementation of Custom Encoding Module*

```
import base64

# The encode module abstracts the native base64 library. This provides the freedom to use whatever
encoding is desired
# and to change it on the fly. Additionally, code is written to allow for the encoding and decoding of long
values.
# This is specifically designed to support encoding of signature from the RSA feature of the PyCrypto
library.
# Both base64 and base32 are supported. Base32 is supported specifically for the purpose of spelling out
words in
# mining algorithms. It is computationally infeasible to spell out more than two letters in mined blocks in
base64;
# however, four and five letter words can be mined with base32 due to the drastically smaller search
space.
# Base64 strings are prepended with "0x40_", which is 64 in hex, and base32 strings are prepended with
"0x20_", which is
# 32 in hex. This aids debugging of encoded strings.


# Encodes input data as base64 and prepends it with "0x40_". "0x40" is 64 in hex and is used to
determines the
# difference between base32 encoded strings and base64 encoded strings in debugging.
# data must be the binary data to be encoded
def encode(data):
    # Pack integers as a byte string
    # https://stackoverflow.com/questions/14764237/how-to-encode-a-long-in-base64-in-python
    if isinstance(data, (int, long)):
        b = bytearray()
        while data:
            b.append(data & 0xFF)
            data >>= 8
        data = b

    return '0x40_'+base64.standard_b64encode(data)


# Decodes input data as base64 and after stripping the "0x40_" prefix. "0x40" is 64 in hex and is used to
determines the
# difference between base32 encoded strings and base64 encoded strings in debugging.
def decode(data, type='string'):
    data = base64.standard_b64decode(data[5:])

    # Unpack the long
    # https://stackoverflow.com/questions/14764237/how-to-encode-a-long-in-base64-in-python
    if type is 'long' or type is 'int':
        data = bytearray(data)  # in case you're passing in a bytes/str
        data = sum((1 << (bi * 8)) * bb for (bi, bb) in enumerate(data))

    return data


# Encodes input data as base64 and prepends it with "0x20_". "0x20" is 32 in hex and is used to
determines the
```

```
# difference between base32 encoded strings and base64 encoded strings in debugging.
def encode_32(data):
    # Pack integers as a byte string
    # https://stackoverflow.com/questions/14764237/how-to-encode-a-long-in-base64-in-python
    if isinstance(data, (int, long)):
        b = bytearray()
        while data:
            b.append(data & 0xFF)
            data >>= 8
        data = b

    return '0x20_'+base64.b32encode(data).lower()


# Decodes input data as base32 and after stripping the "0x20_" prefix. "0x20" is 32 in hex and is used to
determines the
# difference between base32 encoded strings and base64 encoded strings in debugging.
def decode_32(data, type='string'):
    data = base64.b32decode(data[5:].upper())

    # Unpack the long
    # https://stackoverflow.com/questions/14764237/how-to-encode-a-long-in-base64-in-python
    if type is 'long' or type is 'int':
        data = bytearray(data)  # in case you're passing in a bytes/str
        data = sum((1 << (bi * 8)) * bb for (bi, bb) in enumerate(data))

    return data
```

Full Repository: https://github.com/jake-billings/research-blockchain

*Python Implementation of Image Classification Analysis*

```
def process_image_set(filenames, images, type, bottom_threshold=500, top_threshold=9999999):
    imin = -1
    imax = 0

    correct_count = 0

    for i in range(0, len(filenames)):
        print 'Processing %s...' % filenames[i]
        img = images[i].resize((80, 80))
        score = complexity_score(img)
        interesting = score<top_threshold and score>bottom_threshold
        correct = (type.lower() == 'interesting') == interesting
        if correct:
            correct_count += 1

        print '%s, %s, %s\t\t%s\t(%s)' % (type, filenames[i], score, 'INTERESTING' if interesting else
'NONINTERESTING','correct' if correct else 'incorrect')

        if imin is -1:
            imin = score
        if imin > score:
            imin = score
        if imax < score:
            imax = score

    stats = (imin, imax, imax - imin, float(correct_count)/float(len(images))*100)
    print 'min: %s, max: %s, range: %s, accuracy: %s%%' % stats
    return stats


def main(root="test_images"):
    interesting_filenames = os.listdir("%s/interesting" % root)
    uninteresting_filenames = os.listdir("%s/uninteresting" % root)

    interesting_images = []
    for name in interesting_filenames:
        interesting_images.append(Image.open("%s/interesting/%s" % (root, name)))
    uninteresting_images = []
    for name in uninteresting_filenames:
        uninteresting_images.append(Image.open("%s/uninteresting/%s" % (root, name)))

    interesting_stats = process_image_set(interesting_filenames, interesting_images, 'interesting')
    print '----'
    uninteresting_stats = process_image_set(uninteresting_filenames, uninteresting_images,
'uninteresting')
    print '----'
    print uninteresting_stats[1] - interesting_stats[0]

if __name__ == "__main__":
    main(root="test_images")
```

Full Repository: https://github.com/jake-billings/research-blockchain

*Table A1: Classification of Training and Test Images*

| Filename | Category | Correct Classification | Algorithm Classification | Correct? |
|---|---|---|---|---|
| hong_kong.jpg | TRAINING | INTERESTING | UNINTERESTING | INCORRECT |
| ios.png | TRAINING | INTERESTING | UNINTERESTING | INCORRECT |
| ios_screensho.png | TRAINING | INTERESTING | INTERESTING | CORRECT |
| meme.jpg | TRAINING | INTERESTING | INTERESTING | CORRECT |
| morpheus-meme.jpg | TRAINING | INTERESTING | INTERESTING | CORRECT |
| photography.jpg | TRAINING | INTERESTING | INTERESTING | CORRECT |
| photography_2.jpg | TRAINING | INTERESTING | INTERESTING | CORRECT |
| color_rand_147.jpg | TRAINING | UNINTERESTING | UNINTERESTING | CORRECT |
| color_rand_591.jpg | TRAINING | UNINTERESTING | UNINTERESTING | CORRECT |
| monochrome_rand_147.jpg | TRAINING | UNINTERESTING | UNINTERESTING | CORRECT |
| monochrome_rand_147_2.jpg | TRAINING | UNINTERESTING | UNINTERESTING | CORRECT |
| monochrome_rand_591.jpg | TRAINING | UNINTERESTING | UNINTERESTING | CORRECT |
| red_147.jpg | TRAINING | UNINTERESTING | UNINTERESTING | CORRECT |
| red_18.bmp | TRAINING | UNINTERESTING | UNINTERESTING | CORRECT |
| test_image | TEST | INTERESTING | INTERESTING | CORRECT |
| test_image | TEST | INTERESTING | INTERESTING | CORRECT |
| test_image | TEST | INTERESTING | INTERESTING | CORRECT |
| test_image | TEST | INTERESTING | UNINTERESTING | INCORRECT |
| test_image | TEST | INTERESTING | INTERESTING | CORRECT |
| test_image | TEST | INTERESTING | UNINTERESTING | INCORRECT |
| test_image | TEST | INTERESTING | UNINTERESTING | INCORRECT |
| test_image | TEST | INTERESTING | INTERESTING | CORRECT |
| test_image | TEST | INTERESTING | INTERESTING | CORRECT |
| test_image | TEST | INTERESTING | INTERESTING | CORRECT |
| test_image | TEST | INTERESTING | INTERESTING | CORRECT |
| test_image | TEST | INTERESTING | UNINTERESTING | INCORRECT |
| test_image | TEST | INTERESTING | INTERESTING | CORRECT |
| test_image | TEST | INTERESTING | INTERESTING | CORRECT |
| test_image | TEST | UNINTERESTING | UNINTERESTING | CORRECT |
| test_image | TEST | UNINTERESTING | UNINTERESTING | CORRECT |
| test_image | TEST | UNINTERESTING | UNINTERESTING | CORRECT |
| test_image | TEST | UNINTERESTING | UNINTERESTING | CORRECT |